\documentclass[english,aps,twocolumn,10pt,floatfix,superscriptaddress]{revtex4-2}

\usepackage{amsmath}
\usepackage{amssymb}
\usepackage{wasysym}
\usepackage{graphicx}
\usepackage{babel}
\usepackage{array}
\usepackage{verbatim}
\usepackage[colorlinks=true, pdfstartview=FitV, linkcolor=blue, citecolor=blue, urlcolor=blue]{hyperref} %
\usepackage{xcolor}
\usepackage{setspace}

\def\eg       {{\it e.g.}}

\newcommand{\mr}[1]{\mathrm{#1}}
\newcommand{\unit}[1]{\,\mathrm{#1}}
\newcommand{\um}{\,\mu{\rm m}}

\newcommand{\uT}{\,\mu{\rm T}}

\newcommand{\muB}{\mu_{\rm B}}

\newcommand{\ye}{\gamma}

\newcommand{\ket}[1]{\ensuremath{\left|#1\right\rangle}}
\newcommand{\braket}[2]{\ensuremath{\left\langle#1|#2\right\rangle}}

\newcommand{\Bmin}{B_\mr{min}}
\newcommand{\Bmw}{B_\mr{mw}}
\newcommand{\BW}{\Omega_\text{BW}}

\newcommand{\mS}{m_S}

\newcommand{\phieff}{\phi}

\newcommand{\Dt}{\Delta t}
\newcommand{\dt}{\delta t}

\newcommand{\tpeak}{t_\mr{peak}}

\newcommand{\tmin}{t_\mr{min}}

\newcommand{\snr}{\mathrm{SNR}}

\begin{document}
	
\title{Quantum magnetometry of transient signals with a time resolution of 1.1 nanoseconds}
\author{K. Herb}
\email{science@rashbw.de}
\affiliation{Department of Physics, ETH Z\"urich, Otto-Stern-Weg 1, 8093 Z\"urich, Switzerland.}
\author{L.A. V\"olker}
\affiliation{Department of Physics, ETH Z\"urich, Otto-Stern-Weg 1, 8093 Z\"urich, Switzerland.}
\author{J. M. Abendroth}
\affiliation{Department of Physics, ETH Z\"urich, Otto-Stern-Weg 1, 8093 Z\"urich, Switzerland.}
\author{N. Meinhardt}
\affiliation{Department of Physics, ETH Z\"urich, Otto-Stern-Weg 1, 8093 Z\"urich, Switzerland.}
\author{L. van Schie}
\affiliation{Department of Physics, ETH Z\"urich, Otto-Stern-Weg 1, 8093 Z\"urich, Switzerland.}
\affiliation{Department of Materials, ETH Z\"urich, H\"onggerbergring 64, 8093 Z\"urich, Switzerland.}
\author{P. Gambardella}
\affiliation{Department of Materials, ETH Z\"urich, H\"onggerbergring 64, 8093 Z\"urich, Switzerland.}
\affiliation{Quantum Center, ETH Z\"urich, 8093 Z\"urich, Switzerland.}
\author{C. L. Degen}
\email{degenc@ethz.ch}
\affiliation{Department of Physics, ETH Z\"urich, Otto-Stern-Weg 1, 8093 Z\"urich, Switzerland.}
\affiliation{Quantum Center, ETH Z\"urich, 8093 Z\"urich, Switzerland.}
	
\begin{abstract}
Quantum magnetometers based on spin defects in solids enable sensitive imaging of various magnetic phenomena, such as ferro- and antiferromagnetism, superconductivity, and current-induced fields. Existing protocols primarily focus on static fields or narrow-band dynamical signals, and are optimized for high sensitivity rather than fast time resolution.
Here, we report	detection of fast signal transients, providing a perspective for investigating the rich dynamics of magnetic systems. We experimentally demonstrate our technique using a single nitrogen-vacancy (NV) center magnetometer at room temperature, reaching a best-effort time resolution of 1.1\,ns, an instantaneous bandwidth of 0.9\,GHz, and a time-of-flight precision better than 20\,ps. The time resolution can be extended to the picosecond range by use of on-chip waveguides. At these speeds, NV quantum magnetometers will become competitive with time-resolved synchrotron X-ray techniques. Looking forward, adding fast temporal resolution to the spatial imaging capability further promotes single-spin probes as powerful research tools in spintronics, mesoscopic physics, and nanoscale device metrology.
\end{abstract}
	 
\date{\today}
\maketitle
\newpage

\section*{Introduction}
Two-level systems with long coherence times have found widespread application for sensitive detection of a variety of physical parameters, including magnetic~\cite{balasubramanian08,maze08,degen08apl} and electric fields~\cite{dolde11}, temperature~\cite{acosta10}, pressure~\cite{lesik19,hsieh19}, and derived quantities like electrical currents~\cite{chang17,tetienne17} and resistance~\cite{ariyaratne18}.
While often applied to measure static fields, quantum sensing can be extended to detect transient signals by designing suitable control schemes.  Notably, coherent control sequences permit the recording of noise spectra~\cite{bylander11,schmitt17,glenn18}, lock-in detection of harmonic test signals~\cite{kotler11,boss17}, and the acquisition of arbitrary waveform patterns~\cite{magesan13,cooper14,xu16}.
Although demonstrations of wide bandwidths have been made~\cite{dezanche08,chipaux15apl,zopes19,anderson20}, existing protocols are optimized mainly for high sensitivity rather than high temporal resolution. The ability to detect nanosecond or even sub-nanosecond dynamics would significantly widen the application range of quantum sensors, especially in rapidly evolving fields like spintronics~\cite{manchin19} or semiconductor metrology~\cite{orji18}.

In this work, we demonstrate detection of transient magnetic fields with a time resolution approaching one nanosecond.
Our experiment, carried out with a single NV center in a diamond nanopillar, uses a pump-probe scheme consisting of an electrical start trigger and a delayed spin manipulation and readout.
Fast temporal resolution is enabled by combining a speed-optimized pulse sequence~\cite{herb24theory} with efficient microwave pulse delivery.
We show that undesired side effects of short pulses, such as pulse distortions, non-linear spin driving or spurious excitation of non-resonant transitions can be deconvolved by numerical simulation of spin dynamics in the laboratory frame.
We measure transient magnetic test signals for representative application scenarios, including time-resolved imaging of magnetization reversals and domain wall propagation in magnetic nanostructures, as well as time-of-flight detection of magnetic pulses.

\section*{Results}
\subsection*{Scheme for time-resolved quantum sensing}
\begin{figure}
	\includegraphics[width=0.95\columnwidth]{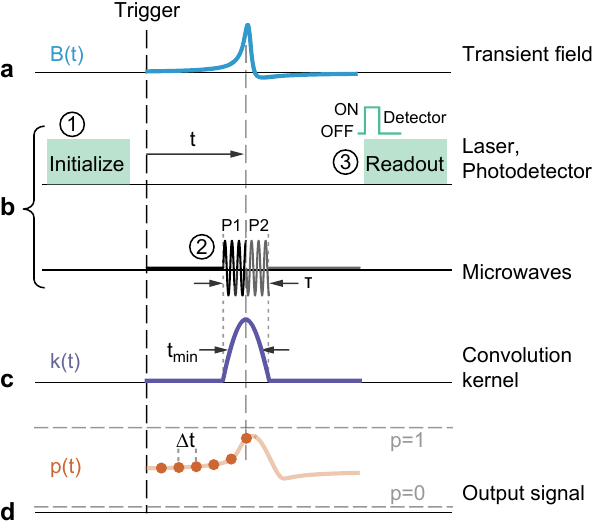}
	\caption{ 
		{\bf Measurement scheme for transient magnetic field sensing.}
		{\bf a}, An electrical trigger initiates play-back of the transient signal $B(t)$.
		{\bf b}, Quantum control sequence for probing the transient signal with the NV spin.  Laser pulses and a fast photo detector are used to initialize (1) and read out (3) the spin state. A composite sequence of two-phase shifted microwave pulses (2, P1-P2), delayed by $t$ with respect to the start trigger, is used to sample $B(t)$ at time $t$.  $\tau$ is the sequence duration.
		{\bf c}, Convolution kernel $k(t)$ of the microwave pulse sequence shown in {\bf b}.  $\tmin$ is the time resolution defined by the full width at half maximum.
		{\bf d}, Measured qubit output signal, given by the state probability $p(t) = |\braket{0}{\psi}|^2$. $p(t)$ is the convolution between $B(t)$ and $k(t)$.
		Each measurement cycle samples one instance of $p(t)$ (dots).  Many repetitions are used to construct the full $p(t)$ trace by equivalent-time sampling. $\Dt$ is the step size.
	}
	\label{fig1}
\end{figure}

Our sensing scheme is based on the generic concept of quantum phase measurements~\cite{degen17}.  The basic idea, illustrated in Fig.~\ref{fig1}, is to reconstruct an unknown transient magnetic signal $B(t)$ by periodically sampling $B(t)$ using a spin probe.  An example of such a transient signal is the magnetic stray field generated by a propagating magnetic domain wall in proximity of the spin probe (Fig.~\ref{fig1}{\bf a}).  To mimic such a measurement, we expose the spin qubit to the influence of a transient field using a short microwave control pulse, leading to a quantum phase pickup $\phieff \propto B(t)$.  This phase pick-up can be read out by a projective measurement.  We then repeat the full measurement cycle and use equivalent-time sampling to reconstruct a trace of the magnetic field transient, equivalent to pump-probe measurements.

The quantum phase measurement is executed in three steps (labeled by 1-2-3 in Fig.~\ref{fig1}{\bf b}). First, we prepare the spin in a known initial state, denoted by $\ket{0}$.  This step is not time critical and may be applied before triggering the transient field.  For the NV center, spin initialization is typically achieved by optical pumping using a green laser pulse~\cite{jelezko04electron}.
Second, the quantum phase measurement is performed.  A sequence of microwave pulses, precisely timed around $t$, temporarily generates a coherent superposition of spin states that interacts with the field $B(t)$.
Third, the resultant final spin state $\ket{\psi}$ is read out via a projective photoluminescence (PL) intensity measurement.  Like the initialization, the read out is not time critical and may be executed once the transient signal has passed.
Together and after many averaging cycles, the PL readout yields a state probability $p = |\braket{0}{\psi}|^2$ that contains an estimate of the desired quantum phase $\phieff$, and thus the field $B(t)$, at time instance $t$ (Fig.~\ref{fig1}{\bf d}).

In the limit of small signals ($|\phieff|\ll \pi/2$, valid for our experiments), the state probability $p(t)$, \textit{i.e.}, the measurable output signal, is proportional to the input transient magnetic field $B(t)$.  The quantitative relation between $p(t)$ and $B(t)$ can be then expressed by a simple convolution, 
\begin{align}
	p(t) = \int_{t'=-\infty}^{\infty} k(t'-t) \ye B(t') \text{d}t' \ .
	\label{eq:pconv}
\end{align}
Here, $\ye$ is the gyromagnetic ratio of the spin probe and $k(t)$ is a convolution kernel (Fig.~\ref{fig1}{\bf c}) that accounts for the details and finite duration of the chosen microwave pulse sequence.  The kernel function is computed using a spin density matrix simulation of the pulse sequence in the laboratory frame (Methods).

To reach a high temporal resolution, the duration $\tau$ of the microwave pulse sequence (Fig.~\ref{fig1}{\bf b}) must be as short as possible.  At the same time, the sequence should maximize $\phieff$ in order to reach adequate sensitivity.  These conflicting requirements necessitate a compromise in measurement speed and acceptable ratio of signal-to-noise (SNR).  In theory, the trade-off between speed and sensitivity is captured by the quantum speed limit (QSL), imposing a bound on how fast a quantum system can be transformed between orthogonal states \cite{mandelstam45}.  In a related theory paper~\cite{herb24theory}, we show that for time-resolved sensing, the QSL can be reached by use of two concatenated microwave pulses of equal amplitude and duration that are phase shifted by $\pi/2$ (P1 and P2 in Fig.~\ref{fig1}{\bf b}). This sequence is equivalent to a Ramsey interferometry measurement with zero time delay~\cite{taylor08}.  The convolution kernel of the composite sequence, plotted in Fig.~\ref{fig1}{\bf c} and derived analytically in \cite{herb24theory}, is given by
\begin{align}
	k(t) =
	\begin{cases}
		\sin[\Omega(\tau/2-|t|)] & |t| < \tau/2 \\
		0                        & |t| > \tau/2 
	\end{cases} \ .
	\label{eq:kernel}
\end{align}
where $\Omega$ is the speed of spin rotations (Rabi frequency) and  $\alpha=\Omega\tau/2$ the spin rotation angle.  The associated time resolution, defined by the full width at half maximum (FWHM) of the kernel function (Fig.~\ref{fig1}{\bf c}), is given by
\begin{align}
	\tmin =  \tau \left(1 - \frac{\arcsin\frac{\sin\alpha}{2}}{\alpha}\right) \ .
	\label{eq:tmin}
\end{align}
Eq.~(\ref{eq:tmin}) confirms our expectation that to achieve a high temporal resolution, fast spin rotations $\Omega \propto \tau^{-1}$ and small rotation angles $\alpha$ are beneficial.

\subsection*{Experimental implementation}

\begin{figure}
  \includegraphics[width=0.95\columnwidth]{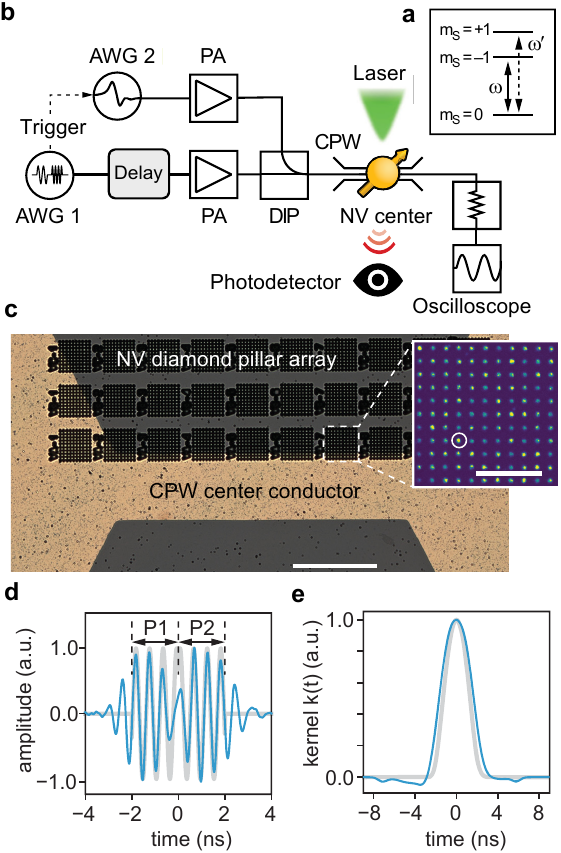}
  \caption{
	{\bf Experimental circuit for NV control and test signal generation.}
	{\bf a}, NV spin energy levels and allowed transitions $\omega$ and $\omega'$.
	{\bf b}, Block diagram of the control circuit.  Microwave pulses and test waveform are generated on separate arbitrary waveform generators (AWG1 and AWG2, respectively), amplified (PA), combined (DIP), and delivered to a common coplanar waveguide (CPW) antenna. A picosecond digital delay generator is used to fine-adjust the relative timing.  See Methods for further details.
	{\bf c}, Optical micrograph of the CPW center conductor (gold) and the diamond nanopillar arrays (black pattern).  The diamond nanopillar array is vertically offset from the CPW by $\sim 25\unit{\um}$.  The inset shows a PL intensity map of a selected pillar array.  High-intensity pillars (one circled) typically contain a single NV center.  Scale bars are $50\unit{\um}$ and $10\unit{\um}$ (inset), respectively.
	{\bf d}, Waveform of a $\tau=4\unit{ns}$ microwave pulse pair.  The gray oscillation is the design waveform and the blue oscillation is the actual waveform measured on the oscilloscope.
	{\bf e}, Corresponding sensing kernels $k(t)$ computed using a spin dynamics simulation (Methods).
  }
  \label{fig2}
\end{figure}

We experimentally demonstrate time-resolved quantum sensing of fast magnetic field transients using a single NV center loated in a diamond nanopillar photonic waveguide~\cite{zhu23}.  The NV center is a $S=1$ spin system with three spin energy levels, $\mS=0$ and $\mS=\pm 1$, and two allowed spin-flip transitions (frequencies $\omega$ and $\omega'$), depicted in Fig.~\ref{fig2}{\bf a}.  To form an effective two-level system, we isolate the $\mS=0$ to $\mS=-1$ transition ($\omega$) by applying an axial bias field of $B_0 = 36\unit{mT}$ along the N-to-V symmetry axis.  For the microwave driving fields accessible in our setup and the chosen bias field, the spin excitation is selective and the $\omega'$ transition can be neglected.  However, for even stronger driving, the full $S=1$ nature of the spin system including the $\omega'$ transition and breakdown of the rotating-wave approximation would have to be taken into account~\cite{kairys23,herb24theory}.

Experimentally, we generate control pulses and magnetic test waveforms on separate circuits and guide them to the NV center via a co-planar waveguide (CPW) antenna (Fig.~\ref{fig2}{\bf b}).  The CPW is fabricated by patterning a gold center conductor onto a quartz coverslip that is narrowed down to $20\unit{\um}$ in the core section to concentrate the microwave current and achieve high microwave field amplitudes (Fig.~\ref{fig2}{\bf c}).  We use a broad-band ($\sim 8\unit{GHz}$) design to allow for short pulse lengths.  Further, we note that precise impedance matching and low absorptive losses are crucial to achieve high Rabi frequencies.

To maximize the Rabi frequency, the NV center pillar array being used (dashed white box in Fig.~\ref{fig2}{\bf c}) is positioned within $\sim 25\unit{\um}$ of the center conductor.  The highest $\Bmw$ amplitudes realized in this work are $\sim 6.3\unit{mT}$, corresponding to Rabi frequencies of approximately $\Omega = \ye\Bmw/\sqrt{2} \sim 2\pi\times 125\unit{MHz}$.

Because control and test pulses have very short durations, on the order of a few nanoseconds, it is crucial to account for pulse distortions in the excitation path~\cite{gustavsson13,chan92}.  To correct for the finite rise and ringdown times of microwave control pulses, we record the pulse waveform on an ancillary oscilloscope connected to the antenna output (Fig.~\ref{fig2}{\bf b}).  Pulse imperfections can then be accounted for by computing the actual kernel function from the measured pulse shape (Methods) followed by a numerical deconvolution of the signal $p(t)$ (see below).  Favorably, even if the actual waveform deviates significantly from the desired waveform (Fig.~\ref{fig2}{\bf d}), the two kernels remain very similar (Fig.~\ref{fig2}{\bf e}).  The dominant effect is a slight broadening of the kernel.
For the test signal, we apply pre-distortion to the digital input waveform until the oscilloscope record matches the desired waveform pattern.

\vspace{-1em}
\subsection*{Detection of transient signals}

\begin{figure*}
	\includegraphics[width=0.7\textwidth]{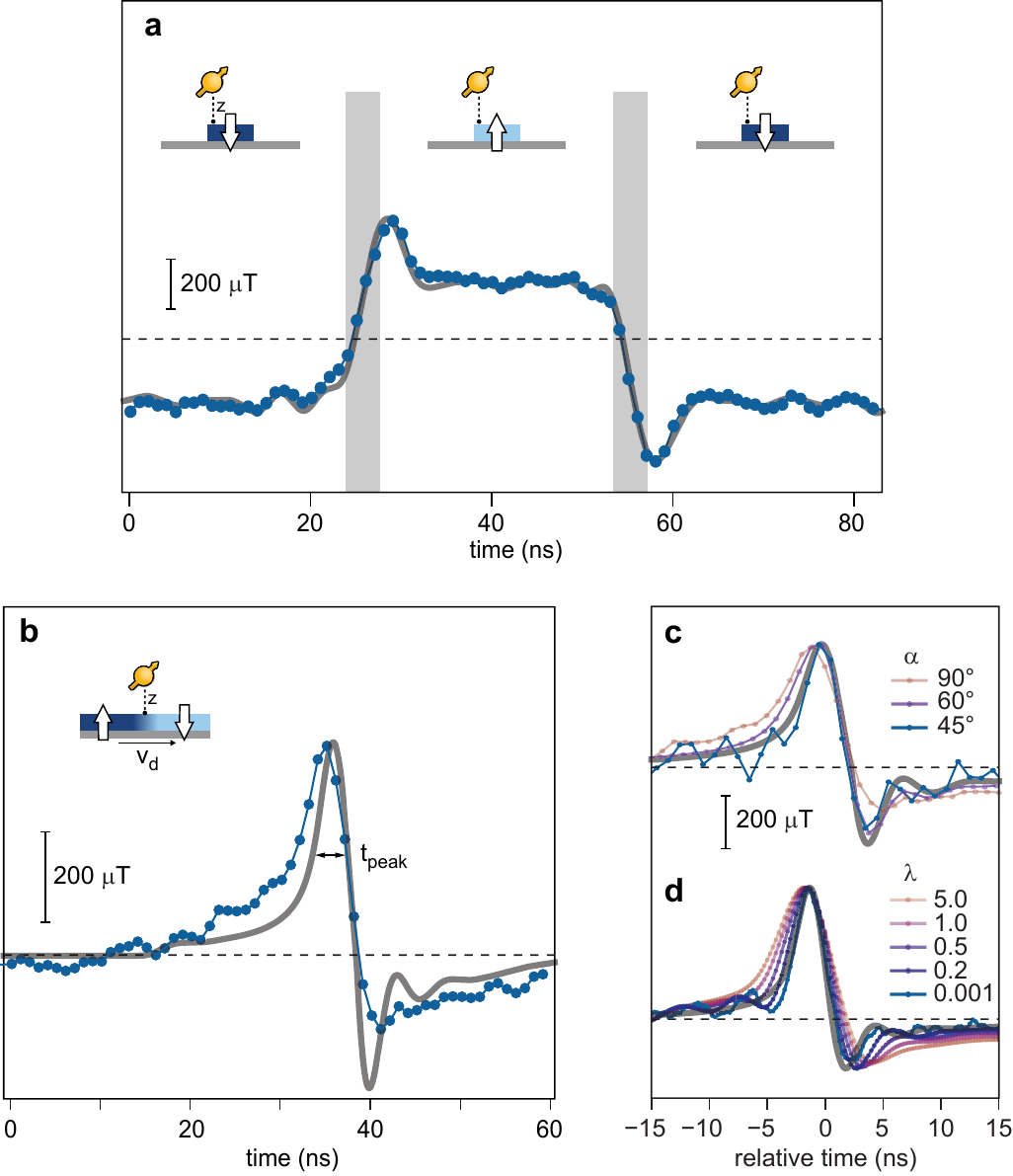}
	\caption{
		{\bf Experimental demonstration of transient magnetic field sensing.}
		{\bf a}, Scenario for detecting current-induced magnetization reversals in a magnetic microdot~\cite{baumgartner17}.  Two reversals are simulated with the magnetization configurations (up or down, indicated by an arrow) shown as insets.
		The gray curve is the input waveform and blue dots show the measured sensor output signal $p(t)$, averaged over $\sim 10^7$ repetitions per point and converted to units of magnetic field.  The $p(t)$ curve is low-pass filtered at $\tau_\mr{3dB} = 0.8\unit{ns}$.  Shaded vertical bars indicate the duration of the $\sim$3-ns-long switching events.  The total measurement time is ca. $1.5\unit{h}$.
		{\bf b}, Scenario for detecting the stray field spike generated near a passing magnetic domain wall (inset).  The gray curve is the input waveform and the blue dots the measured sensor output.  No filtering is applied.  $\tpeak\sim 3\unit{ns}$ is width of the stray field spike.
		{\bf c}, Improved time resolution through shortening of the rotation angle $\alpha$.  Corresponding pulse durations are $\tau = 4, 3, 2\unit{ns}$.  The gray curve shows the input waveform and the blue curves the experimental data.  The $\alpha=90^\circ$ curve is the same as in {\bf b}.
		{\bf d}, Improved time resolution using signal deconvolution by inverse Wiener filtering (see text). $\lambda$ is the regularization parameter.  The gray curve shows the input waveform and the blue curves the experimental data.
		For {\bf a-d}, $\Omega/2\pi = 125\unit{MHz}$ and $\alpha=90^\circ$, unless noted otherwise.
	}
	\label{fig3}
\end{figure*}

As our first example, we consider the transient magnetic field profile associated with magnetization switching events in magnetic nanostructures.  This scenario is representative of applications in non-volatile memories and logic units in spintronic devices~\cite{manchin19}, and of general interest for studying magnetic phase transitions~\cite{kirilyuk10}.  Deterministic switching of magnetization can, for example, be triggered using current or light pulses~\cite{baumgartner17}.  The magnetization reversals typically occur on a time scale of a few nanoseconds~\cite{baumgartner17,sala21,sala22}, and time-resolved detection relies on Hall-effect~\cite{sala21} and anisotropic magnetoresistance measurements~\cite{hayashi07}, the magneto-optical Kerr effect~\cite{atkinson03,beach05}, or synchrotron X-ray imaging~\cite{baumgartner17}.
Moreover, the switching is highly repeatable~\cite{grimaldi20}, which is a prerequisite for most time-resolved measurement techniques (including ours) and required for memory applications.  Here, sensitive quantum probes with nanoscale spatial resolution could add important spatio-temporal information on the switching dynamics.

Fig.~\ref{fig3}{\bf a} displays the synthetic input waveform and measured magnetic field for two simulated magnetization reversals.  Clearly, the output signal recorded by the NV probe closely follows the applied input waveform.  Key details are well resolved, including the transient peak during the $\sim 3\unit{ns}$-long reversal (gray shading) and the ringing before and after the reversal pulses.  Despite the high time resolution and correspondingly short phase accumulation time, the SNR remains high, with a baseline noise of approximately $30\unit{\uT_{rms}}$.  Overall, the result of Fig.~\ref{fig3}{\bf a} demonstrates the feasibility for investigating the magnetization dynamics on nanosecond time scales.

As a second scenario, we consider the transient magnetic field spike produced by a propagating magnetic domain wall.
This scenario is representative for magnetic racetrack devices~\cite{parkin08,velez19}, where domain wall injection and movement is controlled by an electrical current~\cite{dao19}.
Time-resolved imaging of moving domain walls could, for example, yield information about the dynamic distortion of the wall or local variations in the propagation speed~\cite{malozemoff79,thiaville06,baumgartner17}.  Fig.~\ref{fig3}{\bf b} shows the measured NV signal (dots) together with the simulated input waveform.  The simulation assumes a domain wall velocity of $v_\mr{d}=100\unit{m/s}$ and a distance to the spin probe of $z=100\unit{nm}$ (see inset), leading to a magnetic field spike with typical magnitude $\sim 0.5\unit{mT}$ and a duration $\tpeak$ of a few nanoseconds (Methods).  Again, the NV measurement clearly reproduces the main spike feature of the transient signal.  However, contrary to Fig.~\ref{fig3}{\bf a}, some signal broadening is evident.  This broadening is expected, since the microwave pulse duration ($\tau=4\unit{ns}$) is longer than the magnetic field spike ($\tpeak \sim 3\unit{ns}$).  In Figs.~\ref{fig3}{\bf a,b}, the microwave pulse duration $\tau\sim \pi/\Omega$ is determined by the maximum Rabi frequency available with our setup ($\Omega/2\pi \sim 125\unit{MHz}$).

\vspace{-0.1cm}\subsection*{Time resolution and signal deconvolution}\vspace{-0.1cm}

Next, we discuss two approaches to further optimize the time resolution for a given maximum $\Omega$. Both approaches rely on trading SNR for better temporal resolution.
A first method is to reduce the spin rotation angle $\alpha$ (set to $\alpha=90^\circ$ in Figs.~\ref{fig3}{\bf a,b}), equivalent to a narrowing of the kernel function and hence smaller $\tmin$ [see Eq.~(\ref{eq:tmin})].
Fig.~\ref{fig3}{\bf c} shows signal transients with $\alpha = 45-90^\circ$ recorded at maximum $\Omega/(2\pi) = 125\unit{MHz}$.  Decreasing $\alpha$ leads to a small but noticeable improvement in peak width and for $\alpha = 45^\circ$, the broadening is essentially eliminated.  However, an increase in the noise level is also evident.  The time resolution for the shortest pulse ($\tau=2\unit{ns}$ and $\alpha=45^\circ$) is $\tmin = 1.1\unit{ns}$, and the corresponding instantaneous frequency bandwidth is $\BW \approx \tau^{-1} \approx 0.9\unit{GHz}$~\cite{herb24theory}.  These figures are an order of magnitude faster than previous targeted approaches at fast waveform detection~\cite{zopes19,herb20}, and two orders of magnitude faster than methods based on Walsh and Haar reconstruction~\cite{cooper14,xu16}. (Note that these latter experiments were optimized for sensitivity, not time resolution.)

Alternatively, the time resolution can be optimized in post-processing by numerical deconvolution of the kernel function $k(t)$ using inverse filtering.  This approach has the added advantage of removing artifacts caused by pulse distortions in the control circuit, because the true measured kernel function $k(t)$ can be used as an input (Figs.~\ref{fig2}{\bf d,e}).  Further, deconvolution allows taking into account strong spin driving effects (Methods).

Fig.~\ref{fig3}{\bf d} shows the measured waveform ($\alpha=90^\circ$) together with a series of reconstructed waveforms, using Wiener deconvolution (Methods).  Here, $\lambda$ is a unit-less regularization parameter that controls noise suppression.  For weak filtering ($\lambda<1$), the sharp signal spike is accurately resolved at the expense of overall increased noise.  By contrast, strong filtering ($\lambda>1$) efficiently removes noise but broadens the signal peak.
Comparing Figs.~\ref{fig3}{\bf c} and \ref{fig3}{\bf d}, both optimization methods -- shortening of the rotation angle and inverse filtering -- lead to comparable improvements in peak resolution; however, the inverse filtering approach is more powerful as it can account for experimental imperfections and non-linear spin driving, and is applied in post-processing.
A quantitative discussion of the trade-off between sensitivity and time resolution is given in the Methods section.

\begin{figure}
  \includegraphics[width=0.95\columnwidth]{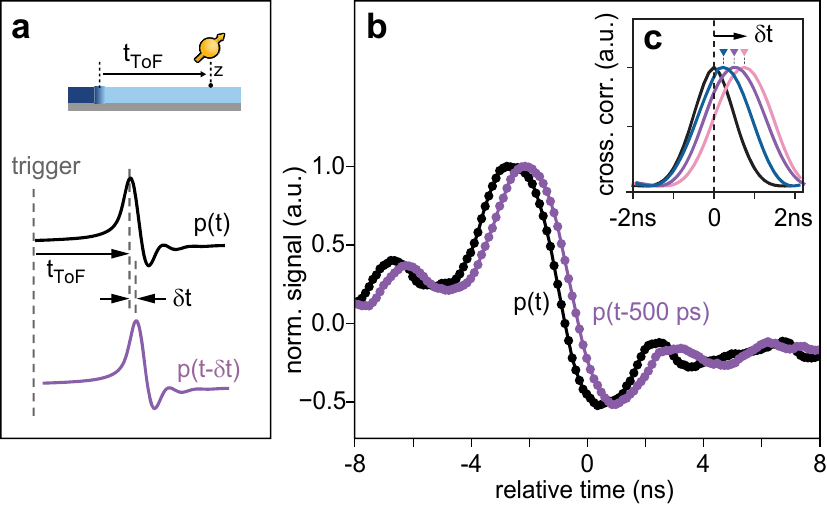}
  \caption{
	{\bf Time-of-flight (ToF) detection.}
{\bf a}, Concept of ToF detection of two time-shifted waveforms, such as magnetic domain walls propagating with different velocities (inset).
{\bf b}, Experimental ToF demonstration for a time shift of $\dt = 500\unit{ps}$.
{\bf c}, Cross-correlated signals $\langle p(t)p(t-\dt)\rangle$ for waveforms with nominal delays of 250, 500 and 750\,ps.  Measured ToF values are determined by the positions of the peak maxima (marked with triangles) and are 249(1)\,ps, 523(2)\,ps and 729(2)\,ps. 
}
  \label{fig4}
  \end{figure}

\subsection*{Time-of-flight detection}

As a further application scenario, we consider time-of-flight (ToF) detection of magnetic pulse events.  In this scenario, rather than mapping the waveform of the transient signal, we are interested in precisely timing the arrival of a magnetic field pulse with respect to a start trigger.  
Such a capability would be useful, for example, to measure the propagation velocity and dispersion of domain walls in magnetic race tracks.
To demonstrate ToF detection, we generate two identical waveforms with a pre-determined delay $\dt$ (Fig.~\ref{fig4}{\bf a}).  Fig.~\ref{fig4}{\bf b} shows recordings of two such waveforms that are time-shifted by $\dt=500\unit{ps}$.  The two transients can be easily distinguished, despite a much slower rise time and overall duration of the waveform.  To quantify the ToF precision, we compute the regularized cross-correlation (via self-deconvolution, see Methods), shown in Fig.~\ref{fig4}{\bf c} for four waveforms with $\dt = 0$, $250$, $500$, and $750\unit{ps}$.  A least-squares fit yields a fit precision of better than $5\unit{ps}$.  The absolute timing error is approximately $\pm 20\unit{ps}$ and limited by jitter in our triggering pulse.
Both numbers could be improved to $\lesssim 1\unit{ps}$, if desired, by using optimized signal averaging and employing hardware with lower jitter.

\clearpage
\section*{Outlook}
In summary, we experimentally demonstrate time-resolved detection of nanosecond magnetic waveforms using the electronic spin of a single NV center in a diamond nanoprobe.  Employing microwave control pulses as short as $2\unit{ns}$, we achieve a best-effort time resolution of $1.1\unit{ns}$, an instantaneous bandwidth of $0.9\unit{GHz}$, and a time-of-flight precision of better than $20\unit{ps}$.  The temporal resolution can be optimized by decreasing the pulse rotation angle or by numerical post-processing.  The method does not make any assumptions on the temporal shape of the transient apart from that the signal must be repeatable and allow synchronization with the NV detection (see below).

Technical upgrades to the microwave delivery should allow improving the time resolution well into the picosecond regime. 
While our demonstration operated at $\Omega/2\pi = 125\unit{MHz}$, on-chip CPWs have enabled Rabi frequencies of $440\unit{MHz}$~\cite{fuchs09}, and nanoscale antenna designs should permit frequencies beyond $1\unit{GHz}$~\cite{rose18}, a roughly 10-fold improvement compared to this study.  At such high frequencies, a number of effects including Bloch-Siegert shifts, violation of the rotating-wave approximation and spurious excitation of the $\omega'$ transition (Fig.~\ref{fig2}{\bf a}) will need to be considered~\cite{bloch40,fuchs09,kairys23,herb24theory}.  Promisingly, these effects are accounted for in the deconvolution step (Methods), because the sensing kernel is calculated by a laboratory-frame simulation of spin dynamics.  Further, NV spin initialization and readout in the $\mS=-1$ or $\mS=+1$ state, rather than the $\mS=0$ state, could be used to mitigate partial excitation of the $\omega'$ transition~\cite{herb24theory}.

Looking forward, combining our technique with scanning probe imaging~\cite{degen08apl,balasubramanian08,casola18,rovny24} will allow studying the nanosecond dynamics of magnetic systems with nanoscale spatial resolution.
Such imaging experiments have recently opened new avenues for analyzing spin textures and spin excitations in nanoscale magnetic systems, ranging from thin-film ferromagnets to ferrimagnets and antiferromagnets~\cite{rondin13,appel19,jenkins19,dovzhenko18,mccullian20,bertelli20,sun21,finco21,velez22}.
Since our method relies on equivalent-time sampling and requires averaging, prerequisites are that measurements can be repeated and triggered. 
This limits the range of systems and dynamics that may be studied.
These restrictions are shared with other time-resolved techniques that rely on pump-probe schemes, including X-ray and magneto-optical imaging, which have been very successful at studying magnetization dynamics.  Examples given include magnetization reversals and domain wall propagation; the former has been demonstrated for up to $10^{11}$ repetitions~\cite{baumgartner17} and with jitter below $0.2\unit{ns}$~\cite{grimaldi20}.
In addition to probing magnetization dynamics, time-resolved sensing also offers the possibility to map time-varying currents allowing, \eg, the \textit{in situ} calibration of rf transmitters~\cite{herb20}, study current dynamics in integrated circuits~\cite{nowodzinski15}, light-induced phase transitions in magnetic materials~\cite{kirilyuk10}, or transient photocurrents in light-sensitive electronic devices~\cite{zhou20}.
Finally, a potential issue of our technique will be perturbation of the system by microwave and optical pulses, which could lead to magnetic, electrical or optical excitation, absorptive heading or thermal drift. Whether such effects are present and how they can be mitigated, for example, by reducing the duty cycle or adjusting antenna design,  will need to be considered on a case-by-case basis.

\section*{Methods}
\subsection*{Experimental apparatus}
Experiments were performed using a home-built confocal microscope equipped with a custom green $\lambda=520\unit{nm}$ diode laser~\cite{welter22thesis} and a single-photon avalanche photo diode (APD, APCM-AQRH, Excelitas).  The laser was directly modulated (on-off) with $\sim 10\,\mathrm{ns}$ resolution.  Arriving photons were separated using a dichroic beamsplitter (Semrock FF526 Di01), time-tagged (National Instruments DAQ 6363) and time-binned in software. 
	
A schematic of the microwave and test signal circuit is shown in Fig.~\ref{fig2}{\bf b}.  Microwave pulses for NV center control were synthesized using IQ modulation of a local oscillator (LO, NI QuickSyn FSW-0020).  The IQ modulation was performed with a home-built IQ modulator circuit based on a HMC1097 chip (Analog Devices).  The intermediate frequency (IF) was provided by the primary arbitrary waveform generator (AWG1, Tektronix AWG5014C) at carrier frequencies between $200-400\unit{MHz}$.
	
Test signals were synthesized using direct sampling on a second waveform generator (AWG2, Keysight M8190A). The AWG offers an instantaneous bandwith of 5 GHz ensuring that test signals requiring high bandwidth can be generated accurately.  For fast demonstration waveforms with frequency components $0.35\unit{GHz}$, we split, in software, the waveform into two streams containing the $<0.35\unit{GHz}$ and $>0.35\unit{GHz}$ frequency components, respectively.  This splitting was necessary because of the frequency bandwidths of the power amplifiers.
AWG1 was further used to synchronize the remaining instrumentation through marker outputs, including modulation of the diode laser, timing of the photon counting, and triggering of AWG2.
	
To fine-adjust the relative timing between AWG1 and AWG2, we either used a programmable delay line (DL1-544-1600PS, MTS Systemtechnik, $5\unit{ps}$ step size) that was interfaced via a home-built interface (ice40HX8K-EVB, Olimex), or the channel skew calibration of AWG1 ($50\unit{ps}$ step size).  This allowed sampling of the magnetic field transient at time increments that were much finer than the timing resolution of AWG1 ($1\unit{ns}$).
	
We used two linear power amplifiers for signal amplification.  The high-frequency channel ($>0.35\unit{GHz}$) containing the microwave pulses and high-frequency stream of the test waveform used a $80\unit{W}$, $0.8-4.2\unit{GHz}$ module (80S1G4, Amplifier Research).  The low-frequency channel ($<0.35\unit{GHz}$) containing the low-frequency stream of the test waveform used a $250\unit{W}$, $0.009-300\unit{MHz}$ module (BBA150, Rohde Schwarz).  On the low-power side, the microwave pulses were combined with the high-frequency stream of the test waveform using a power combiner (ZN2PD2-14W-S+, Mini-Circuits).  On the high-power side, signals were recombined using a custom band diplexer (Universal Microwave Components Corporation).
	
The diplexer output was connected via an ultra-low loss SMA cable (MaxGain 300, Amphenol Times Microwave Systems) to the input of a coplanar waveguide (CPW) acting as the local antenna for exciting the NV center.  The CPW was photo-lithographically defined on a quartz cover slip and placed between the objective and the diamond sample. The CPW had a 3 dB-bandwidth of $\sim 7.5\unit{GHz}$.  The output of the CPW was connected to a high-power attenuator (Meca 697-30-1, 30 dB, $50\Omega$) and further connected to a fast microwave oscilloscope (Rohde Schwarz, RTO2064).  We used the digitized signal of the oscilloscope to correct for imperfections of the pulse and test signal delivery system (see text).
	
The bias field $B_0$ was generated by a NdFeB permanent magnet.  The magnet was mounted on a XYZ translation stage to adjust the magnitude of the bias field and align the field vector along the NV symmetry axis.
		
\subsection*{NV centers}
	
Experiments were performed on electronic-grade diamond single crystals (Element6 Ltd.) with a natural isotope composition.  NV centers were created by $^{15}$N$^+$ ion implantation at an energy of $5\,\mathrm{keV}$ with a fluence of $10^9\unit{cm^{-2}}$, followed by annealing at $1200\unit{^\circ C}$ for 4h in high vacuum and cleaning in a 1:1:1 tri-acid mixture of $\mathrm{H_2SO_4:HClO_4:HNO_3}$ at $120\unit{^\circ C}$.
Nanopillar waveguide arrays were fabricated on membrane samples using electron-beam lithography and RIE etching (QZabre AG) to increase photon yield. The continuous wave (CW) photon count rate of single NV centers was $I_0\sim 1-2\unit{Mcts/s}$ and the optical spin contrast $\epsilon$ between 30\% and 40\%.
	
\subsection*{Wiener deconvolution}
We used Wiener deconvolution to improve the time resolution of our measurements. Wiener deconvolution is closely related to Wiener filtering \cite{press07}. The Wiener filter is a linear filter that minimizes the mean square error between the estimated signal and the original signal. In the situation that the convolution kernel function and the measurement SNR are perfectly known, the Wiener filter is the optimal filter.
	
Consider a noisy signal output $p(t)$, given by
\begin{equation}
	p(t) = k(t) * \ye B(t) + n(t) ,
\end{equation}
where $B(t)$ is the (unknown) transient magnetic field that we want to estimate, $k(t)$ the convolution kernel, and $n(t)$ is the readout noise. Our goal is to construct a filter $c(t)$ such that 
\begin{equation}
	\ye \tilde{B}(t) = c(t) * p(t) \rightarrow \ye B(t)
\end{equation} 
is close to $B(t)$ in the least-squares sense.
In the frequency domain, this is achieved by 
\begin{equation}
	\hat{C}(\omega) = \frac{\hat{K}^\dagger(\omega) |\ye \hat{B}(\omega)|^2}{|\hat{K}(\omega)|^2 |\ye \hat{B}(\omega)|^2 + |\hat{N}(\omega)|^2}.
\end{equation}
where hat symbols denote the Fourier transform of the respective time-domain function.
	
Assuming $\hat{N}(\omega)$ has a flat (white noise) spectrum, we can introduce the dimensionless parameter $\lambda^{-1}=\ye \hat{B}(\omega)/\hat{N}(\omega)$ to parametrize the SNR and obtain an estimate of the input magnetic field,
\begin{align}
	\ye \hat{\tilde B}(\omega) = \frac{ \hat{K}^\dagger(\omega)}{|\hat{K}(\omega)|^2 + \lambda^2} \hat{P}(\omega) .
		\label{eq:wiener_main}
\end{align}
$\lambda$ is adjusted manually and plays the role of a regularization parameter that controls the trade-off between SNR and time resolution.  When decreasing $\lambda$, the deconvolution allows fast features to be reconstructed more accurately while becoming more sensitive to the noise present in a measurement trace.
	
While the Wiener deconvolution can capture linear distortions as produced by pulse distortions, any non-linear effects, such as phase accumulation violating $|\phi|\ll\pi/2$ or the presence of a strong off-axis field component, cannot be reverted.  	For this, the Wiener deconvolution has to be expanded to incorporate a non-linear function. This is known as the Wiener-Hammerstein model. 
	
\subsection*{Kernel function and spin simulations}
To calculate the sensing kernel from an input microwave pulse waveform $B_1(t)$, we performed a density matrix simulation of the spin evolution.  In the experiment, $B_1(t)$ was obtained by direct sampling of the waveform with the downstream oscilloscope, as shown in Fig.~\ref{fig2}{\bf d}.
The simulation was performed in the laboratory frame using the Hamiltonian
\begin{equation}
  	\mathcal{H} = D \hat{S}_z^2  + \ye B_0 \hat{S}_z + \ye B_1(t) \hat{S}_x + \ye B_\mathrm{stim}(t) S_z \ ,
    	\label{eq:hamiltonian}
\end{equation}
where $\ye=28.0345\unit{GHz/T}$ is the gyromagnetic ratio and $D=2\pi\cdot2.87\unit{GHz}$ the zero-field splitting parameter of the NV spin, respectively, and $\hat{S}_x$ and $\hat{S}_z$ are spin-1 matrices.  The first two terms in Eq.~(\ref{eq:hamiltonian}) define the energy spectrum illustrated in Fig.~\ref{fig2}a.  The third term is the oscillating microwave field generated by the microwave antenna; the conversion factor between oscilloscope voltage signal and magnetic field experienced by the NV spin is calibrated using Rabi oscillations.  The fourth term is a $\delta$-like stimulus for sampling the kernel function, implemented as a Gaussian pulse with a line-width much smaller than the (expected) kernel width.
    
The kernel function $k(t)$ was then obtained by time propagating the spin density matrix under the Hamiltonian of Eq.~(\ref{eq:hamiltonian})~\cite{herb22} for a series of stimuli $B_\mathrm{stim}(t-t')$, each value of $t'$ providing one point of the kernel function.  An example for a measured $B_1(t)$ and calculated $k(t)$ is shown in Fig.~\ref{fig2}{\bf d,e}.
    
Importantly, because the simulation is executed in the laboratory frame and involves all three spin states of the NV center ($\mS=0$, $\mS=\pm 1$), the simulation naturally captures excitation of the non-resonant transition and Bloch-Siegert-shifts due to the counter-rotating term associated with $B_1(t)\hat{S_x}$.
	
\subsection*{Calibration of magnetic signal amplitude}
The measured photon counts were converted to transition probabilties $p(t)$ by normalizing them with reference values for $\mS=0$ and $\mS=-1$ recorded concurrently.  To calibrate the signal amplitude, pulsed ODMR measurements were performed where a constant test signal was applied using the diplexer.  The carrier frequency of the $<1\unit{us}$ long $\pi$ pulse was varied. The extracted frequency shift was used to obtain a calibration for the data shown in Figs.~\ref{fig3},\ref{fig4}.

\subsection*{Sensitivity}

\begin{figure}
	\vspace{-0.5em}
	\includegraphics[width=0.95\columnwidth]{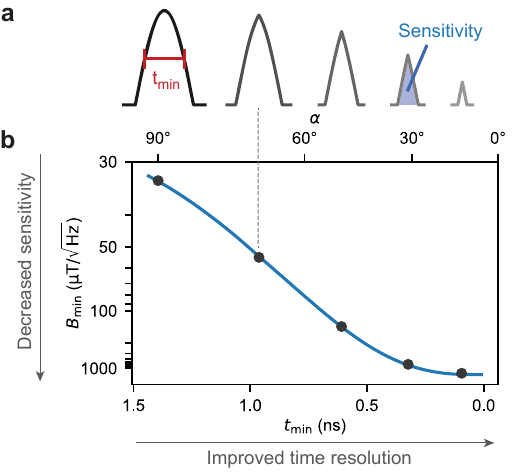}
	\caption{{\bf Trade-off between sensitivity and time resolution.}
		\textbf{a}, Kernel function $k(t)$ for decreasing rotating angles $\alpha$. The time resolution $\tmin$ is given by the FWHM and the sensitivity (given by $\Bmin$) is proportional to the area under $k(t)$.
		\textbf{b}, $\Bmin$ plotted against $\tmin$ for the experimental parameters given with Eq.~(\ref{eq:counts}).}
	\label{fig:sensitvitiy}
	\vspace{-0.5em}
\end{figure}

Starting from Eq. (6) in Ref.~\cite{herb24theory}, we estimate the photon counts $C$ associated with a certain transition probability $p(\phi=\gamma_{e} B \tau)$ by
\begin{equation}
	C(\phi) = \left[1-\epsilon p(\phi)\right] C_0 = \left[1-\epsilon p(\phi)\right] \frac{I_0 T t_\mathrm{int}}{t_\mathrm{seq}} \ ,
	\label{eq:counts}
\end{equation}
where $\epsilon$ is the optical contrast of the NV center, $I_0$ the count rate in continuous-wave illumination, $T$ the total measurement time, $t_\mathrm{int}$ the integration time in photon binning, and $t_\mathrm{seq}$ the duration of the sensing sequence (given by the time between subsequent triggers).
Representative values for our measurements are $I_0 \sim 1-2\unit{Mcts/s}$, $\epsilon = 0.3-0.4$, $t_\mathrm{int} \sim 300-500\unit{ns}$, and $t_\mathrm{seq} \sim 2.5\unit{\mu s}$.  We typically averaged until reaching $C_0 \sim  0.1-1\unit{Mcts}$, corresponding to total measurement times $T \sim 1-10\unit{s} $, 

From Eq.~\ref{eq:counts}, we estimate the shot-noise-limited SNR to
\begin{equation}
\begin{split}
		\snr &= \frac{C(\phi)-C(\phi=0)}{\sqrt{C(\phi=0)}} \\
&= \frac{\epsilon \phi \left(1 - \cos{\left(\alpha \right)}\right) \sin{\left(\alpha \right)}}{\alpha \sqrt{4 - 2 \epsilon \sin^{2}{\left(\alpha \right)}}} \sqrt{C_0} \ .
\end{split}
\end{equation}
From this equation we deduce the minimum detectable field (the ``sensitivity'') by solving $\snr= 1$ for $\Bmin$ using $\phi = \ye\Bmin \tau$. This results in 
\begin{equation}
	\Bmin =  \frac{\alpha \sqrt{4 - 2 \epsilon \sin^{2}{\left(\alpha \right)}}}{\epsilon \gamma_{e} \tau \left(1-\cos{\left(\alpha \right)}\right) \sin{\left(\alpha \right)}} \sqrt{C_0} \ .
	\label{eq:Bmin}
\end{equation}
For a rotation angle of $\alpha=\pi/2$, a kernel duration of $\tau=2\unit{ns}$ and the above measurement parameters, we find a nominal sensitivity of $\Bmin\sim 35\unit{\mu T/\sqrt{Hz}}$.
To inspect how sensitivity can be traded for improved time resolution, we plot $\Bmin$ [given by Eq.~(\ref{eq:Bmin})] against $\tmin$ [given by the FWHM of the sensing kernel, Eq.~(\ref{eq:tmin})], shown in Fig.~\ref{fig:sensitvitiy}. 

\subsection*{Time-of-flight measurements}
To estimate the time-of-flight (ToF) between two sampled waveforms, we needed to estimate the time delay. Numerous methods exist in the field of signal processing for this purpose. Here, we choose an approach that does not require precise \textit{a priori} knowledge of the sensing kernel or waveform shape.  In an ideal, free-space scenario, a time lag $\dt$ can be seen as a convolution with a kernel $k(t)=\delta(t-\dt)$.  Therefore, by deconvolving one measured signal with the other, one obtains the shared convolution kernel. For an acquisition with finite sampling, this can be modeled by a sinc function and fitted. By using the Wiener deconvolution method, we can again regularize the solution allowing for a solution optimal for the given signal-to-noise ratio.
	
\subsection*{Magnetic test waveforms}
	
\textit{Domain wall propagation} --
The simulation assumed a magnetic domain wall in a thin magnetic film with a surface magnetization of $25\unit{\muB/nm^2}$.  The time trace was calculated by first computing the spatial stray field profile $B(x)$ across a domain wall extending along $y$ and located at $x=0$, and then converting the spatial profile into a temporal profile by setting $t=x/v_\mr{d}$, where $v_\mr{d}=100\unit{m/s}$ is the domain wall velocity.  Here, $B(x)$ is the vector component projected onto the NV symmetry axis, given by $\theta=54^\circ$ and $\phi=0^\circ$, where $\theta$ is the polar angle and $\phi$ the azimuth, respectively.  The stand-off distance was assumed to be $z=150\unit{nm}$. 
	
\textit{Magnetization reversal of magnetic disk} --
The simulation assumed a disk of $1\unit{\um}$ diameter and a surface magnetization of $75\unit{\muB/nm^2}$.  The reversal was simulated by a domain wall that passed across the disk~\cite{baumgartner17} with a velocity of $100\unit{m/s}$.  The domain wall was assumed to have a width of $50\unit{nm}$ and a left N\'eel chirality; however, smaller domain walls and other chiralities do not much affect the magnetic stray field.  The NV center was positioned at $z=100\unit{nm}$ above the center of the disk.  Its vector orientation was along $\theta=54^\circ$, $\phi=90^\circ$. Note that real world signals such as the magnetization reversal in magnetic dots might be affected by jittering in the order of a few nano seconds \cite{sala21,grimaldi20}, which might induce further blurring. For these devices, jitter could for example be reduced by higher current densities or inplane magnetic fields or adjusting the device geometry.  However, this limitation is inherent to all measurement approaches that cannot capture the transient in a single shot.

\bibliography{library_kh}

\vspace{1em}

\section*{Acknowledgements}
The authors thanks Hugo Karras and Prof. Gunnar Jeschke for providing the Keysight M8190A AWG for the measurements and their help during setup, and Dr. Theodor Prosch for fruitful discussions. This work has been supported by the SNSF through Project Grant No. 212051, and by the SBFI through Project QMetMuFuSP" under Grant. No. UeM019-8, 215927.  J.M.A. and L.A.V. acknowledge funding from SNSF Project Grant No. 201590.

\section*{Author contributions}
K.H. and C.L.D. conceived the experiment with input from N.M., L.v.S. and P.G..  K.H. carried out the experiments with the help of L.A.V.. J.M.A. assisted in the preparation of diamond samples. K.H. and C.L.D. performed the data analysis. All authors discussed the results.

\section*{Competing interests}
The authors declare no competing interests.

\section*{Data availability}
The data that support the findings of this work are available from the corresponding authors upon reasonable request.

\end{document}